# Clustering Drives Cooperation on Reputation Networks, All Else Fixed

Tamas David-Barrett

Email: tamas.david-barrett@trinity.ox.ac.uk

Address: Trinity College, Broad Street, Oxford, OX1 3BH, UK

Web: www.tamasdavidbarrett.com

**Abstract**

Reputation-based cooperation on social networks offers a causal mechanism between graph properties and social trust. Recent papers on the 'structural microfoundations' of the society used this insight to show how demographic processes, such as falling fertility, urbanisation, and migration, can alter the logic of human societies. This paper demonstrates the underlying mechanism in a way that is accessible to scientists not specialising in networks. Additionally, the paper shows that, when the size and degree of the network is fixed (i.e., all graphs have the same number of agents, who all have the same number of connections), it is the clustering coefficient that drives differences in how cooperative social networks are.

Keywords: social networks, evolution of cooperation, reputation, free-riding clustering coefficient, agent-based, structural microfoundations.

# Introduction

Early work on the evolutionary conundrum of costly cooperation focused on interactions that were dyadic in nature: the behaviour was seen as taking place between two entities. Why would Being A help Being B if it is costly for A and benefits only B? This question is dyadic as it concerns the interaction only between A and B. Alternatively, the question could be posed for interactions between an individual and a group: why would Being C give up something precious for a group D, to which C may belong but so do many others? In this case as well, the interaction is dyadic, it is between C and D, even if the latter is an entity on a higher organisational level.

Two vastly different solutions to this puzzle were provided. First, the inclusive fitness mechanism showed that cooperation can emerge among close relatives based on shared DNA [1-6]. Second, the reciprocal interaction mechanism showed that cooperation can emerge if agents have repeated interactions, can remember each other, and can adjust their behaviour according to the past actions of the other [7-10]. This latter mechanism also had evolutionary foundations and was present in a wide range of species apart from humans [11].

Even though these two solutions were entirely different, one following the logic of biology, the other of economics, they were identical in that they both lacked interaction structure. Bacterium A helped bacterium B in a dyadic fashion, and hero C gave up its resources for the group D in a one-to-many way, also dyadic between individual and a 'blob' of a group. While the idea that the interactions could form a network was present, the fact that the structure of the network could be important for the rise of cooperation had been ignored. It was ignored, because in a network in which the question is about *dyadic* cooperation, it can be interesting if one interacts with few or many, but it is irrelevant whether those others are also connected to each other and if so how.

Two insights changed the approach to cooperation.

First, the assumption that all interesting processes necessarily take place dyadically was altered by the anthropological, and also everyday, observation that people gossip [12-14]. When they do this, they pass on information about shared acquaintances, behind the acquaintance's back. And for this, at least three agents are needed. Being A tells Being B that their shared connection C did something wrong, for instance, by adopting a cheating rather than a cooperative stance. Gossip like this is an excellent way of speeding up the detection of cheaters, and thus it is a robust way of ensuring cooperation [15-22].

Second, network science, which emerged in interaction with, but following a different historical logic to the problem of cooperation, provided a method to structure reputation dynamics [23-25]. The network science approach highlighted the importance of the social network's structural properties and pointed out that the cooperation-enhancing effect of reputation increases with higher interconnectedness [23, 26, 27].

Thanks to the successful merger of these two traditions, the past 20 years has seen the emergence of a large literature that looks at the interaction between characteristics of a social network, in particular, the density and the degree distribution, and the space for costly cooperation to emerge [28-37]. Empirical findings from experiments on humans in several cultures has provided evidence that clustering alone does not, or at least does not necessarily, promote cooperation; it can, however, be a powerful driver when combined with a space for reputation formation [22, 38-41].

It is to this tradition that the current paper aims to offer a small contribution. Recent papers suggested a causal link between macro-societal demographic processes and the social network's micro structure, showing how falling fertility, urbanisation, and migration can reduce the propensity to cooperate among individuals [42, 43]. These models relied on the literature's insights on the relationship between network structure and cooperation but lacked the instantiation for the case of $k$-regular $n$-sized connected graphs. Furthermore, many scientists, who are not in the field of network science but are interested in the consequences of the 'structural microfoundations' models, asked for an explanation of the underlying mechanics. The objective of this paper is to fill this gap.

## Methods and Results

The first part of methods introduces a game in which the agents meet repeatedly, but their interactions have no structure, and there is no possibility for gossip. The second part introduces graph structure and gossip into the same repeated game framework. The third part shows how the local interconnectedness, measured in the clustering coefficient, affects cooperation.

NB. The primary aim of this paper is to illustrate the mechanics of the relationship between the clustering coefficient and cooperation for scientists who are not themselves in the field of repeated games on networks. The models are particular instantiations of the literature's findings, which first appeared stated differently but with the same qualitative meaning, three decades ago [23-26], and has been supported by both theoretical and empirical evidence since [22, 27-41, 44-47].

## Everyone plays everyone, nobody gossips

Let us define a social group made of $n$ agents. The agents play dyadic prisoner's dilemma games where the payoff matrix is symmetric, and is

$$p = \{\{p_{1,1}, p_{1,2}\}, \{p_{2,1}, p_{2,2}\}\}$$

Let us assume that $0 < m < n$ players are randomly chosen as having type 'cheater', and thus $n\text{-}m$ players have type 'cooperator'.

Each player tracks every other player's type, where $d_{i,j}$ is i's agent's expectation of j's type. Initially all agents assume that everyone else is a co-operator.

Agents are randomly paired, and play their strategy the following way: if agent $i$ expect agent $j$ to be a cooperator, then agent $i$ plays her strategy according to her type. If, however, she thinks that $j$ will be a cheater, then she will cheat independent of her own type. In other words, if we expect the other to play nice, we will play nice if we are nice, and play dirty if we are naughty, but if we expect the other to play dirty, then we always play dirty even if we are nice. Formally:

$$\{i,j\} \sim U\{1, \ldots, n\} \mid i \neq j$$

that is, we randomly pick the two agents. Then

$$a_i = \begin{cases} \text{cheat if } d_{i,j} = \text{cheater} \\ \text{type}_i \text{ if } d_{i,j} = \text{cooperator} \end{cases}$$

where $a_i$ is the action of agent $i$.

After their interaction, the agents update their expectation of the other's type to the action played by the other:

$$d_{i,j} = a_j$$
$$d_{j,i} = a_i$$

Notice that because of the way the actions are chosen, if a partner sees you as cheater even just once, they will never trust you again.

Let the interactions repeat until an average agent is in plays $r$ times. The repeat number, $r$, and the number of cheaters, $m$, drive the total payoffs (Fig.1a).

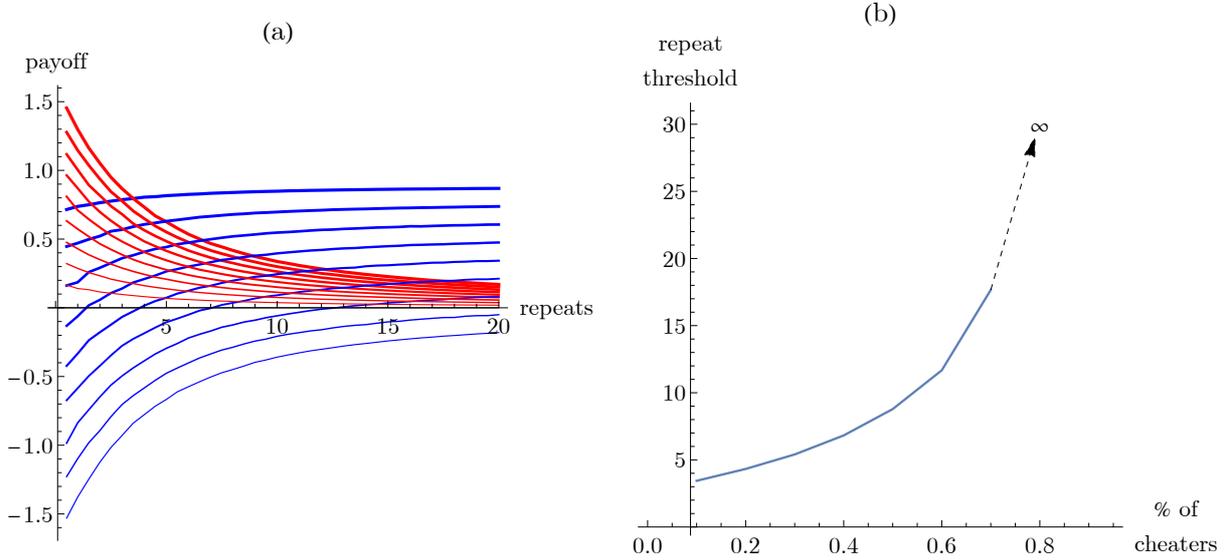

Fig 1. The interaction between repeat number, payoffs, and cooperation thresholds. Panel (a) shows the cooperators' and cheaters' average payoffs as a function of the number of repeated interaction per agent. Blue: cooperator, red: cheater. The thickness of the lines corresponds to the proportion of cooperators among the agents, with the thickest (topmost) lines corresponding the case when there is a only a single cheater, i.e., m=1, and the thinnest (bottommost) corresponding to the case when there is only a single cooperator, i..e, m=9. Panel (b) shows how many repeated interactions are needed for the cooperating strategy to have a higher payoff than the cheating one, as a function of the proportion of cheaters. (Average of 1000 repeats. NB. Note that the line always goes to infinite if n-m=1 and the payoffs are prisoner dilemma structure. That is, this model is not about the emergence of cooperation, only the maintenance. Payoffs $p=\{\{1,-1.6\},\{1.5,0\}\}$, see SM1.)

The results (Fig.1a) show that

- Cheater payoff decreases and cooperator payoff increases as the number of interactions increase independent of the number of cheaters (That is, the red lines are all going down, and the blue lines are all going up.)
- As the cheater number increases the payoffs of both cheaters and cooperators increase. (The thinner lines are under the bolder ones, for both colours.)
- With the increasing cheater number, the payoff drops faster for the cooperators than the cheaters. (The blue lines take a larger space than the red ones.)
- As a consequence, the lines corresponding to the same number of cheaters (the red and blue lines of the same thickness) meet at different repeat numbers.

The crossing point of the red and the blue lines, given a particular number of cheaters, determines the repeat number above which it is advantageous to be a cooperator. The higher, the cheater number, cooperation repeat threshold is also higher (Fig. 1b). Notice that the fact that the line in Fig. 1b is increasing means that once cooperation emerges an evolutionary or learning mechanism would turn the entire group into cooperators.

## Social network and gossip

Let us change the above 'plain vanilla' model by introducing a fixed interaction structure in terms of a network, and allow gossip.

Let us assume that the $n$ agents form a connected graph of degree $k$. (That is, each dot has $k$ 'friends'.) I use this structure because it most closely resembles a human group in which $n$ is magnitudes larger than $k$. Our species' groups tend to be so large that no individual can meaningfully be connected to every other group member, while the advantage of having more connections pushes people to have more connection if they can [48]. So, people tend to end up with about a similar range of friends.

Let us also assume that each time an agent is in an interaction, she decides her move based on what she thinks about the alter's type *and* what her trusted friends think, where her 'trusted' friends are those agents that are connected to her, and she still thinks that their type is cooperator.

$$e_{i,j} = \begin{cases} \text{cheater if } d_{i,j} = cheater \\ \text{cheater if } d_{i,j} = cooperator \text{ and } \dfrac{\tilde{k}}{(k+1)} \geq 0.5 \\ \text{cooperator otherwise} \end{cases}$$

Where $\tilde{k}$ is the number of trusted friends who think that $j$ is a cheater.

Let us set the action of $i$ the based on this expectation, similar to above:

$$a_i = \begin{cases} \text{cheat if } e_{i,j} = cheater \\ type_i \text{ if } e_{i,j} = cooperator \end{cases}$$

On the surface, the results are very similar to the previous (Fig. 2).

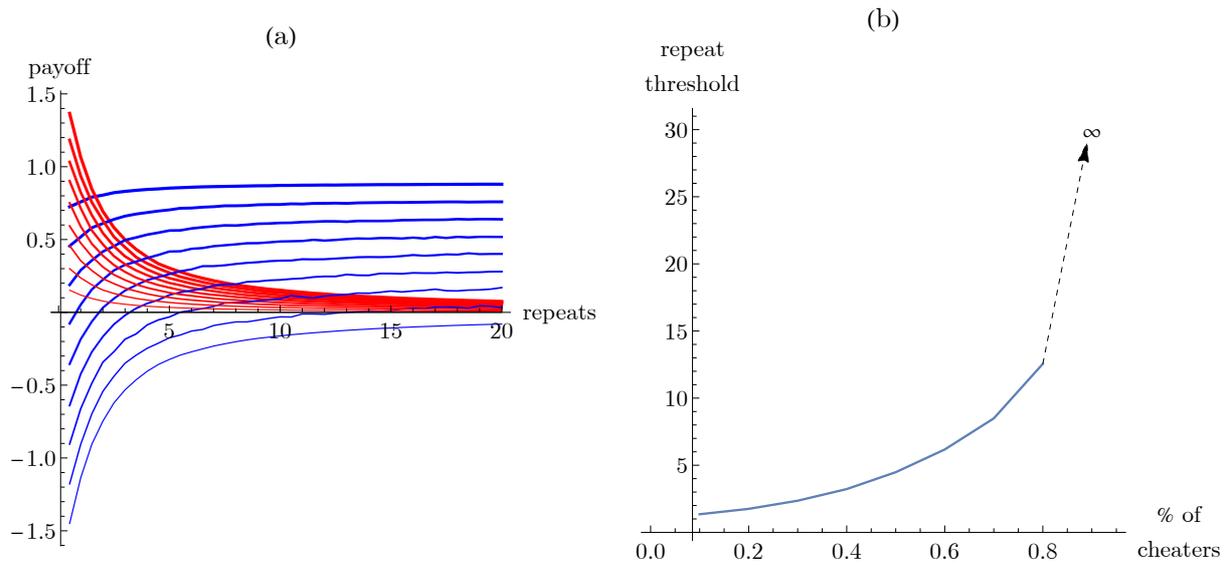

Fig. 2. The interaction between repeat number, payoffs, and cooperation thresholds when the agents form a network and gossip. (For the definitions of the curves, see Fig. 1. Average of 1000 simulations. For each repeat, the network was randomly generated, with parameters $n=10$, and $k=4$.)

Comparing the two results illustrates why gossip matters for cooperation: faster flow of information reduces the interaction repeat threshold (Fig. 3.)

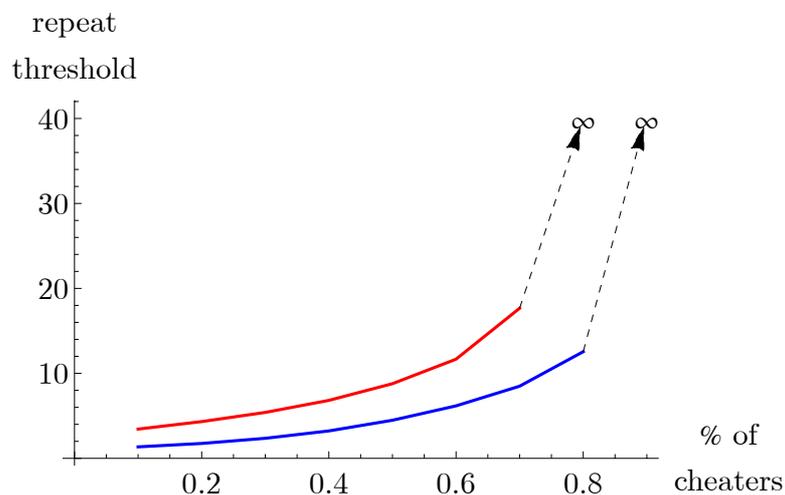

Fig. 3. Cooperation threshold comparison between the 'plain vanilla', and 'gossip-on-network' cases. The red curve is identical to Fig. 1b, while the blue curve is the same as Fig. 2b.

Thus, gossip allows the identification of cheaters, and thus the shift to a cooperative group.

## The importance of network structure

Notice that in the gossip-on-network version above, the key to the improved cheater detection was that people who were connected to each other were able to pass on information, i.e., they gossiped and tracked others' reputation. By the nature of gossip as defined here, it only matters if there is a *shared* connection between the two agents exchanging information about the third. As they are only interacting with those that are in their social network, such gossip can only take place if there is a closed network-triangle among them.

Given that we set the parameters at $n=10$, and $k=4$, it is known that there are 59 different non-isomorphic graph structures, that is, they are different to each whichever direction you turn them [49]. There is considerable variation among them in terms of number of closed triangles (Fig. 3).

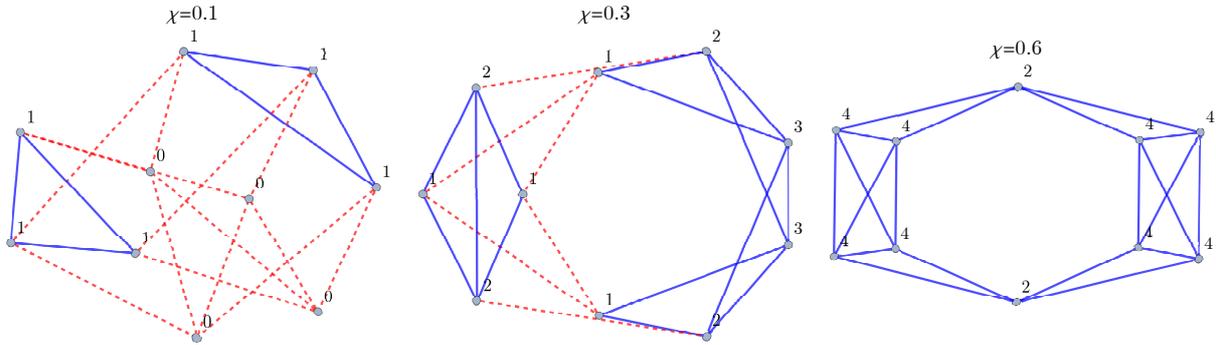

Fig. 4. Illustration for variation in clustering coefficient. (All three graphs are 4-regular 10-sized: they have ten nodes, with each of the nodes having 4 connections. The number next to each node is the number of closed triangles that node has, and $\chi$ is the average clustering coefficient. Blue lines: edges in closed triangles, red dotted lines: edges that are not part of a closed triangle.)

The average clustering coefficient, $\chi$, of this set of 59 graphs ranges from 0% to 70%. (The clustering coefficient is the proportion of closed triangles compared to maximum possible triangles. That is, it is a measure of how interconnected the agents are, with both the size of the group and the number of 'friends' staying the same.)

I calculated the cooperation threshold for each graph in the set, for the example of $m=1$. The results show that all else equal, the clustering coefficient drives the differences in cooperation among the graphs (Fig. 5). The downward sloping line shows that the higher the clustering coefficient is, the easier it is for cooperation to emerge.

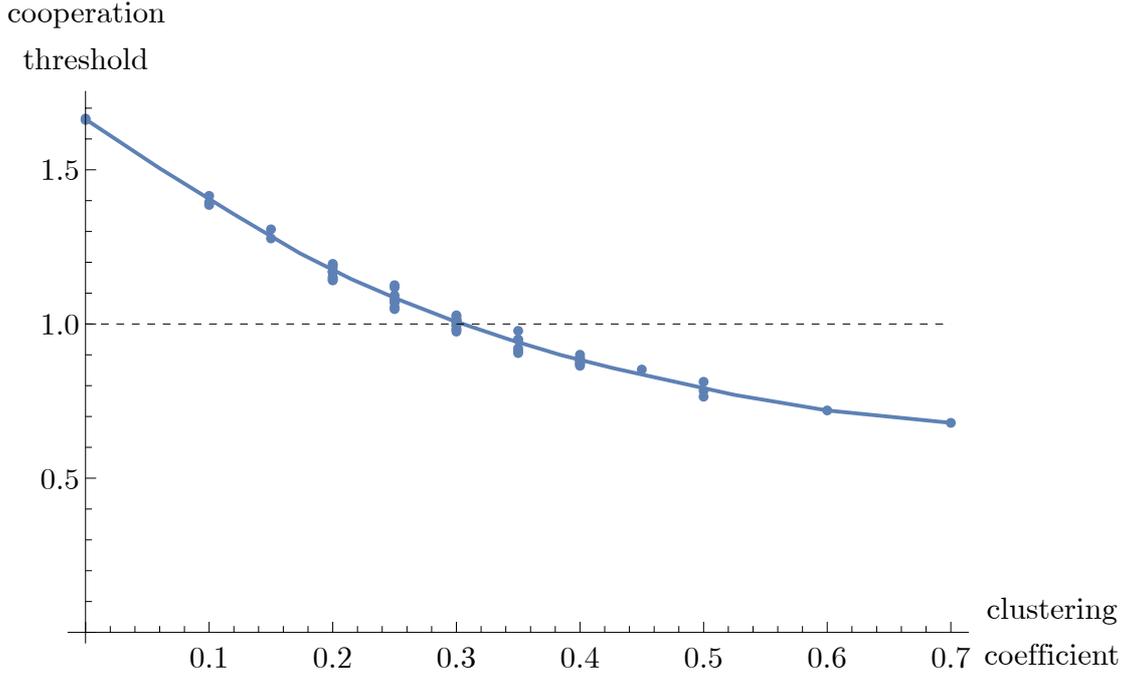

Fig. 5. The clustering coefficient drives the differences in how easily cooperation emerges on graphs that all have 10 nodes, with each node having 4 connections. Each blue dot represents a different graph, the dashed line is a regressed interpolation. (x-axis: average clustering coefficient of the graph, y-axis: threshold for average number of interactions above which the cooperative strategy is more advantageous than cheating. Number of cheaters: 1, randomly chosen in each new simulation, repeat number: 1000.)

Notice that the threshold curve is not only downward sloping in Fig. 5, but it is also under 1 when $\chi$ is higher than about 30%. That means that in this particular example gossip can spread so fast that cheating becomes a loss-making strategy even before the cheater meets every one of its connections. (For cases in which there are more than one cheater, i.e., $m>1$, see SM 2.)

# Discussion

The role of the clustering coefficient is central to 'structural microfoundations' models [42, 43]. In these, one or more demographic processes (such as, falling fertility, urbanisation, migration, deadly wars and epidemics) reduce the clustering coefficient of the social network, which in turn triggers adaptive individual responses (such as, increased norm violations, the rise of homophily-based friendship, a shift to value fundamentalism, and increased susceptibility to type-signalling institutions) and social responses (such as, the rise of law, the emergence of kin-cue-using ideologies, and the fake news industry). At the heart of these models is the assumption that, all else being equal, there is a causal relationship between the clustering coefficient and the level of cooperation in human societies.

Although the main goal of this paper is to provide an illustration for the underpinnings of the 'structural microfoundations' models, by showing how the clustering mechanics of social trust works in the particular case of the $n$-sized $k$-regular connected graphs, it also introduces a small contribution to the literature. The effect of the clustering coefficient on cooperation stance in this particular graph family has not been spelled out before (to my knowledge).

The main finding, i.e., that the higher the clustering coefficient the more cooperation, has some limitations. Reaching the maximum clustering coefficient may not be optimal for collective action. This might be the case, for instance, when there are parallel behaviours [50], like teaching the 'information value' [51, 52]. NB. This observation has parallels with some of our earlier work on the negative effects of social stratification, in which the loss of collective action efficiency came from a particular status dynamics [53].

I can see two immediate future extensions.

The first has to do with the optimal way to formulate expectations about the future behaviour of others. There are many ways to form perceptions about reputation and respond to them, and the particular choice of reputation assessment method interacts with the chance and speed of the rise of cooperation [54]. It would be interesting to ask if the optimal reputation assessment technique is dependent on the clustering coefficient, and thus likely to shift with demographic processes.

Second, it would be intriguing to ask if some of the vast kaleidoscope of social technologies that human cultures invented in history, are ways to increase the clustering coefficient. In particular, do institutions that regulate inequality, a social technology that is present in all human societies, increase social network interconnectedness?

# Supplementary Material

**Supplementary Material 1: Rationale for choosing the particular payoff matrix**

The payoff matrix I chose is quantitively different from the traditional payoffs used in the prisoners' dilemma game, although not qualitatively. The reason is that this paper is to illustrate the benefits from cooperation, when there is a small payoff from cheating on others.

The justification for $p=\{\{1,-1.6\},\{1.5,0\}\}$ is as follows.

- When both agents cooperate, they both receive a unit resource, i.e., 1.
- This is not a zero-sum game. That is, there is synergy in cooperation, which is lost in asymmetric interaction. Thus, when one cheats, and the other cooperates, the cheater takes only part of the total benefit from cooperation, i.e., less than 2. At the same time, it pays to cheat, thus the cheater receives more from cheating than from cooperation, i.e., more than 1. Otherwise, we would not have a cooperation dilemma at all. Thus, from these two inequalities we have the cheater's payoff between 1 and 2, hence the choice of 1.5.
- There is friction in life: being cheated on is a little more costly than the gain of the cheater. Hence -1.6 is the payoff the cooperator receives when the other cheats.
- The baseline is no cooperation. Hence when they both cheat, then nothing happens, they both receive 0.

Of course, as long as the game is set up as a cooperation dilemma, any parameters within that constraint would yield the same qualitive results.)

**Supplementary Material 2: Clustering effect on cooperation threshold when *m*>1**

Gossip is a social technology that allows tracking reputation using via third-party information [15-22, 43]. Thus, the point of the gossip mechanism is that it increases the speed of cheater detection. This suggests that the more cheaters there are, the less important the gossip is, and thus the less important the clustering coefficient is. The simulation results are consistent with this observation (Fig. S1).

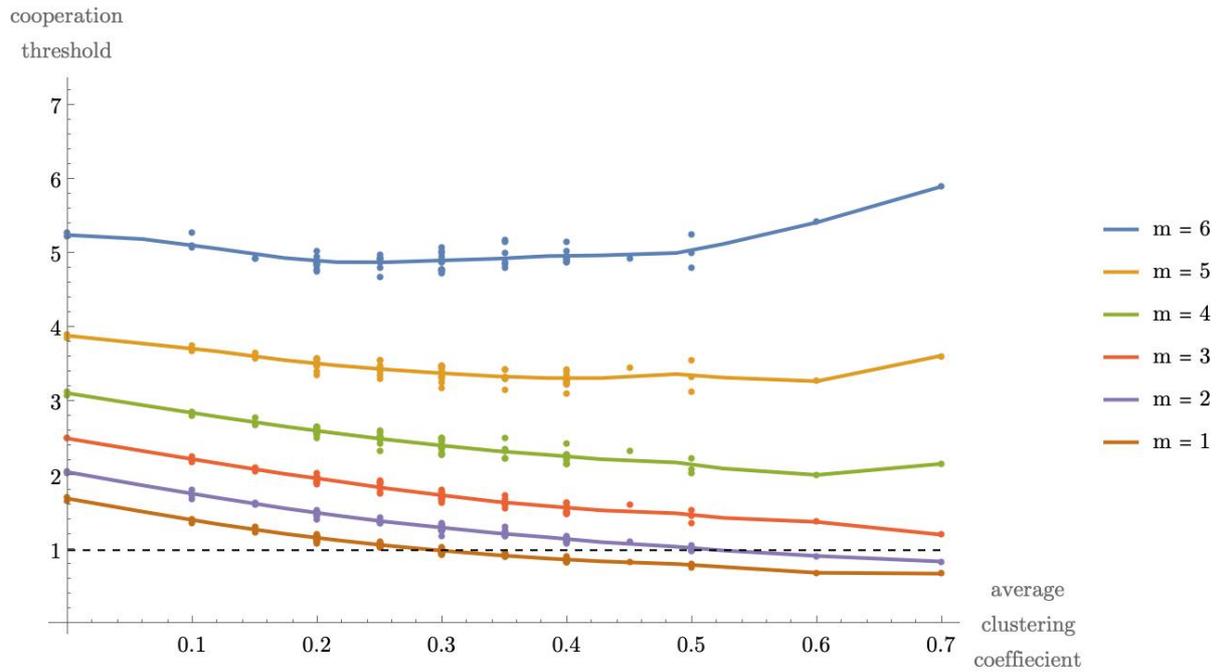

Fig. S1. Clustering coefficient's effect on cooperation wanes as cheater number increases. (Each dot represents one element of the 4-regular, 10-sized, connected graphs' set, which has 59 elements. x-axis: average clustering coefficient of the graph, y-axis: threshold for average number of interactions above which the cooperative strategy is more advantageous than cheating. Number of cheaters: m, randomly chosen in each new simulation, repeat number: 1000.)